\documentclass[a4paper,twocolumn,english,aps,prb,floatfix,showpacs,preprintnumbers]{revtex4}
\usepackage[latin9]{inputenc}
\usepackage{color}
\usepackage{amsmath}
\usepackage{amssymb}
\usepackage{graphicx}
\usepackage{esint}

\makeatletter

\@ifundefined{textcolor}{}
{%
 \definecolor{BLACK}{gray}{0}
 \definecolor{WHITE}{gray}{1}
 \definecolor{RED}{rgb}{1,0,0}
 \definecolor{GREEN}{rgb}{0,1,0}
 \definecolor{BLUE}{rgb}{0,0,1}
 \definecolor{CYAN}{cmyk}{1,0,0,0}
 \definecolor{MAGENTA}{cmyk}{0,1,0,0}
 \definecolor{YELLOW}{cmyk}{0,0,1,0}
}

\usepackage{babel}

\usepackage{babel}

\usepackage{babel}

\usepackage{babel}

\makeatother

\usepackage{babel}
\begin{document}

\title{Conductivity of epitaxial and CVD graphene with correlated line defects }

\author{T. M. Radchenko,$^{1}$ A. A. Shylau,$^{2}$ and I. V. Zozoulenko$^{3}$}

\affiliation{$^{1}$Deptartment of Solid State Theory, Institute for Metal Physics
of NASU, 36 Acad. Vernadsky Blvd., UA-03680 Kyiv, Ukraine}

\affiliation{$^{2}$Center for Nanostructured Graphene (CNG), Department of Micro-
and Nanotechnology, DTU Nanotech, Technical University of Denmark,
DK-2800 Kongens Lyngby, Denmark}

\affiliation{$^{3}$Laboratory of Organic Electronics, ITN, Linköping University,
SE-60174 Norrköping, Sweden}

\date{\today }
\begin{abstract}
Transport properties of single-layer graphene with correlated one-dimensional
defects are studied using the time-dependent real-space Kubo--Greenwood
formalism. Such defects are present in epitaxial graphene, comprising
atomic terraces and steps due to the substrate morphology, and in
polycrystalline chemically-vapor-deposited (CVD) graphene due to the
grain boundaries, composed of a periodic array of dislocations, or
quasi-periodic nanoripples originated from the metal substrate. The
extended line defects are described by the long-range Lorentzian-type
scattering potential. The dc conductivity is calculated numerically
for different cases of distribution of line defects. This includes
a random (uncorrelated) and a correlated distribution with a prevailing
direction in the orientation of lines. The anisotropy of the conductivity
along and across the line defects is revealed, which agrees with experimental
measurements for epitaxial graphene grown on SiC. We performed a detailed
study of the conductivity for different defect correlations, introducing
the correlation angle $\alpha_{\textrm{max}}$ (i.e. the maximum possible
angle between any two lines). We find that for a given electron density,
the relative enhancement of the conductivity for the case of fully
correlated line defects in comparison to the case of uncorrelated
ones is larger for a higher defect density. Finally, we study the
conductivity of realistic samples where both extended line defects
as well as point-like scatterers such as adatoms and charged impurities
are presented. 
\end{abstract}


\maketitle

\section{Introduction}

As for all crystalline solids, the presence of a certain amount of
disorder in graphene,---the thinnest known material nicknamed as a
``miracle material'' due to its superior properties,\cite{Novoselov2012}---is
dictated by the second law of thermodynamics. Defects, playing a role
of disorder, are always present in graphene samples due to the imperfection
of the fabrication processes, and even can be not always stationary,
migrating with a certain mobility governed by the activation barrier
and temperature.\cite{Krasheninnikov2011} Such migration and relaxation
to the equilibrium state as well as the features of the growth technology
can result in a correlation in the configuration of point or/and line
defects. The effect of the spatial correlations of point defects on
the transport properties of graphene is currently under debate. A
recent observation of the temperature enhancement of the conductivity
of exfoliated graphene was attributed to the effect of dopant correlations.\cite{Fuhrer2011}
This conclusion was based on the semi-classical predictions relying
on the standard Boltzmann approach within the Born approximation.\cite{Sarma2009}
At the same time, numerical calculations within the time-dependent
real-space Kubo--Greenwood formalism showed\cite{Radchenko 1} that
correlation in the spatial distribution of short- and long-ranged
point defects do not lead to any enhancement of the conductivity in
comparison to the uncorrelated case.

In epitaxial graphene the surface steps caused by substrate morphology
are spatially correlated and act as line scatterers for the charge
carriers.\cite{Kuramochi2012} Epitaxial graphene films grown on SiC\cite{Kuramochi2012,Held2012}
(by SiC decomposition) or on Ru\cite{Gunther2011} (by CVD method)
comprise two distinct self-organized periodic regions of terrace and
step, leading to ordered graphene domains.\cite{Gunther2011} Step
edges, as well as single defects, can be visualized by the Kelvin
probe force microscopy (KPFM),\cite{Melitz2011,Held2012} the scanning
tunneling microscopy (STM) or the atomic force microscopy (AFM). For
instance, a width of the steps, observed from the AFM images, remained
nearly constant, about 10 nm;\cite{Kuramochi2012} while the step
heights identified from the KPFM and the STM varied respectively from
0.09 up to 0.75 nm\cite{Held2012} and from 0.5 up to 1.5 nm.\cite{Ji2011}
Scanning tunneling potentiometry (STP) measurements of a local electric
potential (as current flows through a graphene film) demonstrate\cite{Ji2011}
that local perturbations caused by the substrate atomic steps are
critical to transport in graphene. The STP potential measurements
are now possible not only on micro- and macroscopic length scales,\cite{Ji2011}
but on mesoscopic scale as well,\cite{Wang2012} where the quantum
nature of transport manifests itself directly.

Experimental measurements show an increase of the resistance with
the step density,\cite{Dimitrakopoulos2001} the step heights,\cite{Ji2011}
and the step bunching.\cite{Lin2011} Also, an anisotropy of the conductivity
in the parallel and perpendicular directions to the steps is revealed,
which is due to higher defect abundance in the step regions.\cite{Kuramochi2012,Yakes2010}
Substrate steps alone increase the resistivity in several times relative
to a perfect terrace,\cite{Ji2011} with the ratio of the estimated
electron mobilities in the terrace and step regions being about 10:1.\cite{Kuramochi2012}
Despite the strong curvature of graphene in the vicinity of steps,
a structural deformation contributes only little to electron scattering.\cite{Low2012}
For the SiC substrate, the dominant scattering mechanism is provided
by the sharp potential variations in the vicinity of the step due
to the electrostatic doping from the substrate strongly coupled with
graphene in the step regions.\cite{Low2012}

The anisotropic charge transport have been also revealed in CVD-grown
graphene due to the parallel orientation of the quasi-periodic nanoripples.\cite{Ni2012}
The charged line defects are believed to represent the limiting scattering
mechanism of the electronic mobility in CVD graphene.\cite{FerreiraEPL,Radchenko 2,Tuan2013,Ihnat}
In this case the line defects correspond to grain boundaries separating
grains of different (or the same) crystal orientations\cite{Yazyev2010NatMat}
or quasi-periodic nanoripples originated from the metal substrate.\cite{Ni2012}
In contrast to the grain boundaries in the CVD-grown graphene exhibiting
mainly (but not always)\cite{Ni2012} a random network of lines, the
substrate atomic steps acting as the line defects in epitaxial graphene
manifest correlation in their orientation and even can be almost parallel
to each other as a result of epitaxial growth.\cite{Kuramochi2012,Gunther2011,Held2012}
Taking into account that the charged line defects govern the electrical
transport in CVD-grown graphene, the atomic-stepped line defects in
epitaxial graphene are also expected to affect strongly its transport
properties and even govern them especially if the correlation can
be controlled.

The main aim of this paper is therefore to investigate the influence
of the orientational correlation of the extended line defects on the
conductivity of graphene. We do this numerically, utilizing the quantum
mechanical time-dependent real-space Kubo method\cite{Radchenko 1,Roche_SSC,Yuan10,Leconte11,Markussen,Ishii,Lherbier12,Tuan2013}
allowing us to study graphene sheets approaching the realistic dimensions
of millions of atoms.

The paper is organized as follows. In Sec. II, the basic model of
the system at hand including a model potential for one-dimensional
(1D) defects and the basics of the numerical Kubo method are formulated.
Section III presents and discusses the obtained numerical results.
Finally, the conclusions of our work are given in Sec. IV.

\section{Tight-binding model and Kubo--Greenwood formalism}

\begin{figure*}
\includegraphics[width=1\textwidth]{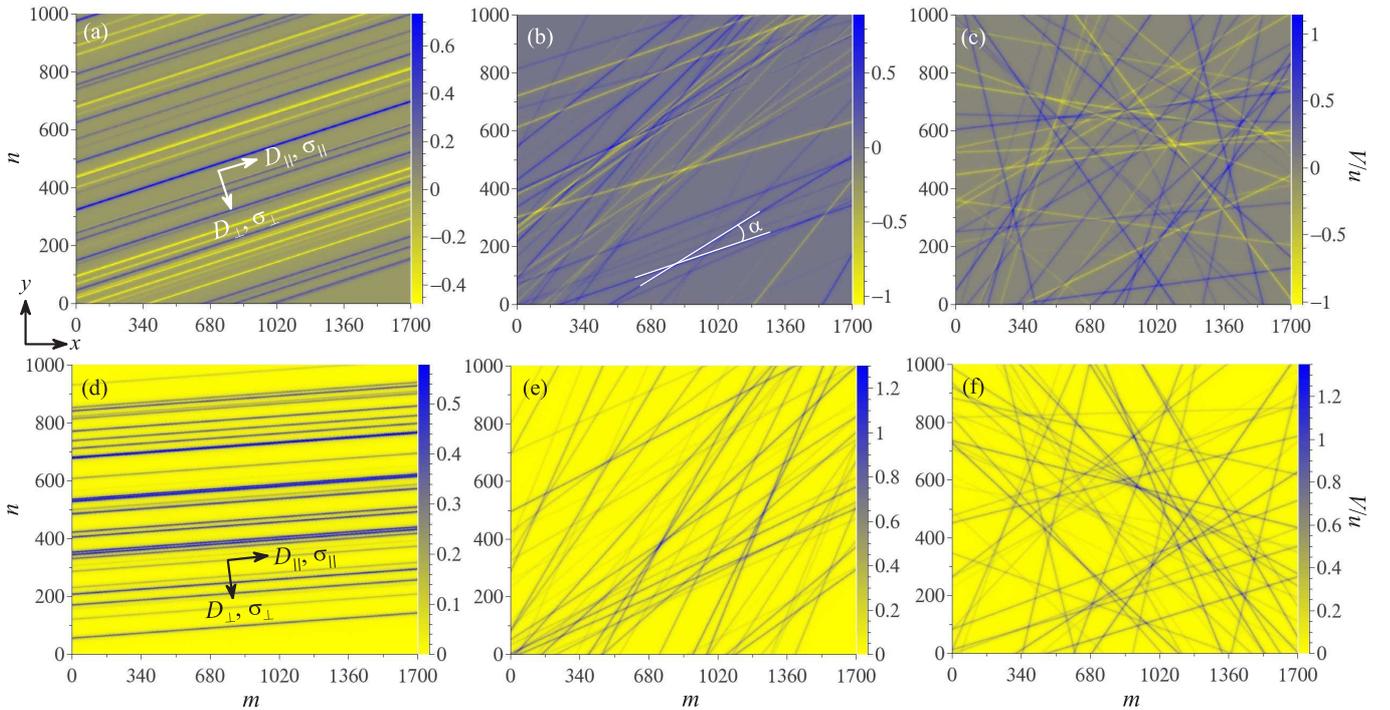}

\caption{(Color online) Distributions of (a)--(c) symmetric, $V\gtrless0$,
and (d)--(f) asymmetric, $V>0$, scattering potentials, Eq. (\ref{Eq_Lorentzian}),
for a representative configuration of 50 orientationally correlated
line defects with different correlation angles $\alpha_{\textrm{max}}$
(the maximal possible angle between any two lines): $0^{\circ}$ (a),
(d); $45{}^{\circ}$ (b), (e); $90^{\circ}$ (c), (f). Note, that
$\alpha_{\textrm{max}}=0^{\circ}$ and $\alpha_{\textrm{max}}=90^{\circ}$
correspond to the cases of parallel and random (totally uncorrelated)
lines, respectively. The graphene lattice size is $m\times n=1700\times1000$
sites corresponding to $210\times210$ nm. The maximum potential height
$\triangle=0.25u$.}

\label{Fig_Potential} 
\end{figure*}

We model electron dynamics in graphene using a standard $p$-orbital
nearest-neighbor tight-binding Hamiltonian defined on a honeycomb
lattice,\cite{Castro Neto review,Peres_review,DasSarma_review}

\begin{equation}
\hat{H}=-u\sum_{i,i^{\prime}}c_{i}^{\dagger}c_{i^{\prime}}+\sum_{i}V_{i}c_{i}^{\dagger}c_{i},\label{Eq_Hamiltonian}
\end{equation}
where $c_{i}^{\dagger}$ and $c_{i}$ are the standard creation and
annihilation operators acting on a quasiparticle on the site $i$,
the summation is carried out over all nearest-neighbor sites $i$
and $i^{\prime}$ of graphene lattice, $u=2.7$ eV is the hopping
integral for the neighboring C atoms $i$ and $i^{\prime}$ with the
distance $a=0.142$ nm between them, and $V_{i}$ is the on-site potential
describing scattering by defects. For the case of epitaxial graphene
on SiC(0001) the scattering is related to an electrostatic doping
at the steps from the interface state on the Si-terminated substrate
which has a metallic character with a high density of interface states.\cite{Low2012,Mattausch,Varchon}
Considering this state as a charged line following the step, the effective
potential for electron in graphene can be calculated in an analytical
form using the Thomas--Fermi (TF) approximation. \cite{FerreiraEPL,Radchenko 2}
The effective TF potential is expressed via the cosine and sine integral
functions, and can be well fitted by the Lorentzian-shaped function.\cite{Radchenko 2}
If there are $N_{l}$ 1D scatterers in graphene, we model them by
the effective long-range Lorentzian-type potential, 
\begin{equation}
V_{i}=\sum_{j=1}^{N_{l}}U_{j}\frac{A}{B+x_{ij}^{2}},\label{Eq_Lorentzian}
\end{equation}
where $U_{j}$ is a potential height, $x_{ij}$ is a distance between
the site $i$ and the line $j$, and the fitting parameters $A$ and
$B$ depend on the considered charge carrier (electron) densities.\cite{Radchenko 2}
In the present study, we consider two cases, namely, symmetric (attractive
and repulsive, $V\gtrless0$), and asymmetric (repulsive for electrons,
$V>0$) scattering potentials, where $U_{j}$ are chosen randomly
in the ranges $\left[-\Delta,\Delta\right]$ and $\left[0,\Delta\right]$,
respectively, with $\Delta$ being the maximum potential height. Figure
\ref{Fig_Potential} illustrates the potential shape for both symmetric
($V\gtrless0$) and asymmetric ($V>0$) cases, where, to characterize
the relative positions of lines, we introduced correlation angle $\alpha_{\textrm{max}}$---the
maximum possible angle between any two lines. Note that an electrostatic
doping discussed above represents one of possible mechanisms leading
to the enhanced scattering at the steps in epitaxial graphene. Other
sources of scattering can include, for example, trapped silicon atoms
that tend to aggregate at the step edges.\cite{Yakes2010} We therefore
consider our scattering potential, Eq. (\ref{Eq_Lorentzian}), as
a phenomenological model which with a proper adjustment of the parameters
can describe both short- and long-range scattering caused by the presence
of 1D defects.

To calculate numerically the dc conductivity $\sigma$ of epitaxially-
or CVD-grown graphene layers with 1D extended defects, the real-space
order-$N$ numerical implementation within the Kubo--Greenwood formalism
is employed, where $\sigma$ is extracted from the temporal dynamics
of a wave packet governed by the time-dependent Schrödinger equation.\cite{Radchenko 1,Roche_SSC,Yuan10,Leconte11,Markussen,Ishii,Lherbier12}
This is a computationally efficient method scaling with the number
of atoms in the system $N$, and thus allowing treating very large
graphene sheets containing many millions of C atoms.

A central quantity in the Kubo--Greenwood approach is the mean quadratic
spreading of the wave packet along the $x$-direction at the energy
$E$, $\Delta\hat{X}^{2}(E,t)=\bigl\langle\hat{(X}(t)-\hat{X}(0))^{2}\bigr\rangle$,
where $\hat{X}(t)=\hat{U}^{\dagger}(t)\hat{X}\hat{U}(t)$ is the position
operator in the Heisenberg representation, and $\hat{U}(t)=e^{-i\hat{H}t/\hbar}$
is the time-evolution operator. Starting from the Kubo--Greenwood
formula for the dc conductivity,\cite{Madelung}

\begin{equation}
\sigma_{xx}=\frac{2\hbar e^{2}\pi}{\Omega}\text{Tr}[\hat{v}_{x}\delta(E-\hat{H})\hat{v}_{x}\delta(E-\hat{H})],\label{Eq_Kubo-Greenwood}
\end{equation}
where $\hat{v}_{x}$ is the $x$-component of the velocity operator,
$E$ is the Fermi energy, $\Omega$ is the area of the graphene sheet,
and factor 2 accounts for the spin degeneracy, the conductivity can
then be expressed as the Einstein relation, 
\begin{equation}
\sigma\equiv\sigma_{xx}=e^{2}\tilde{\rho}(E)\lim_{t\rightarrow\infty}D(E,t),\label{Eq_sigma(t)}
\end{equation}
where $\tilde{\rho}(E)=\rho/\Omega=\textrm{Tr}[\delta(E-\hat{H})]/\Omega$
is the density of sates (DOS) per unit area (per spin), and the time-dependent
diffusion coefficient $D(E,t)$ is related to $\Delta\hat{X}^{2}(E,t),$
\begin{align}
D(E,t)= & \frac{\bigl\langle\Delta\hat{X}^{2}(E,t)\bigr\rangle}{t}\notag\\
= & \frac{1}{t}\frac{\text{Tr}[\hat{(X}_{H}(t)-\hat{X}(0))^{2}\delta(E-\hat{H})]}{\text{Tr}[\delta(E-\hat{H})]}.\label{Eq_Diffusion}
\end{align}

It should be noted that in the present study we are interested in
the diffusive transport regime, when the diffusion coefficient reaches
its maximum. Therefore, following Refs.~\onlinecite{Leconte11,
Lherbier12}, we replace in Eq. (\ref{Eq_sigma(t)}) $\lim_{t\rightarrow\infty}D(E,t)\rightarrow D_{\max}(E)$,
such that the dc conductivity is defined as 
\begin{equation}
\sigma=e^{2}\tilde{\rho}(E)D_{\max}(E).\label{Eq_sigmaMax}
\end{equation}
The DOS is also used to calculate the electron density as $n(E)=\int_{-\infty}^{E}\tilde{\rho}(E)dE-n_{\text{ions}},$
where $n_{\text{ions}}=3.9\cdot10^{15}$ cm$^{-2}$ is the density
of the positive ions in the graphene lattice compensating the negative
charge of the $p$-electrons {[}note that for an ideal graphene lattice
at the neutrality point $n(E)=0${]}. Combining the calculated $n(E)$
with $\sigma(E)$ given by Eq. (\ref{Eq_sigmaMax}) we compute the
density dependence of the conductivity $\sigma=\sigma(n)$. Details
of numerical calculations of DOS, $D(E,t)$, and $\sigma$ are given
in Ref.~\onlinecite{Radchenko 1}.

\section{Results and Discussion}

\begin{figure*}
\includegraphics[width=1\textwidth]{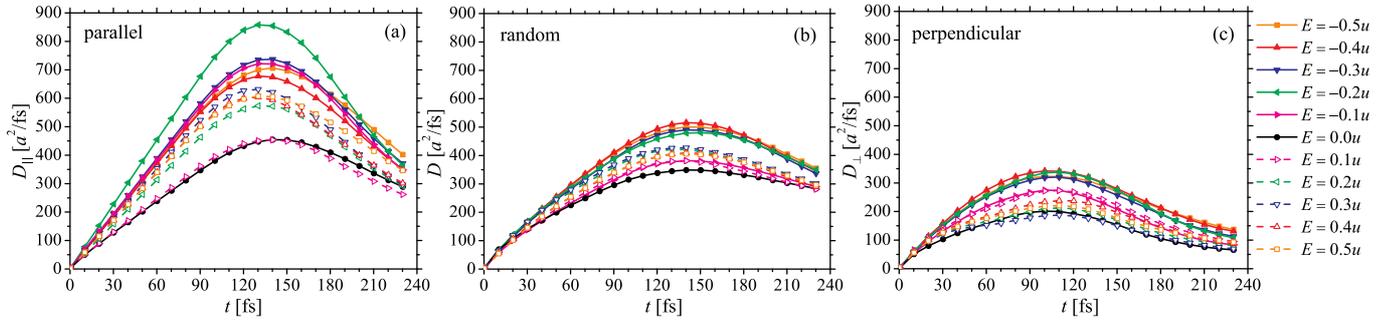}

\caption{(Color online) Time-dependent diffusion coefficients within the energy
interval $E\in[-0.5u,0.5u]$ for 50 line defects, which are (a) parallel
$(D_{\Vert})$, (b) randomly distributed $(D\equiv D_{\mathrm{rnd}})$,
or (c) perpendicular $(D_{\bot})$ to the direction of transport (i.e.
\textit{x}-direction) (see Fig. \ref{Fig_Potential}). $V\gtrless0$,
$\Delta=0.25u$.}

\label{Fig_Diffusivity} 
\end{figure*}

This section contains numerical results for the dc conductivity calculated
using the time-dependent real space Kubo--Greenwood formalism within
the tight-binding model briefly presented in the previous section.
We perform calculations for three different concentrations of 1D defects
corresponding to 10, 50, and 100 line defects in the $1700\times1000$-size
lattice (see Fig. \ref{Fig_Potential}). Parameters $A$ and $B$
entering the scattering potential are weakly electron-density dependent.
By fitting the analytical expression for the Thomas--Fermi potential
to the Lorentzian-type one, Eq. (\ref{Eq_Lorentzian}), \cite{Radchenko 2}
in the range of representative electron densities $1\cdot10^{-5}\lesssim|n_{e}^{\exp}|\lesssim5\cdot10^{-5}$
atom\texttt{$^{-1}$} used in the present study, we obtain $A=33.57$
and $B=16.96$. All calculations for both symmetric, $V\gtrless0$,
and asymmetric, $V>0$, potentials are performed for the potential
strength $\Delta=0.25u=0.675$ eV, which is close to the values of
the contact potential variation at the substrate atomic steps observed
in epitaxial graphene by means of the Kelvin probe force microscopy.
\cite{Held2012,Ji2011,Wang2012}

Figure \ref{Fig_Diffusivity} shows the time evolution of the diffusion
coefficient, Eq. (\ref{Eq_Diffusion}), within the energy interval
$E\in[-0.5u,0.5u]$ for the symmetric potential, $V\gtrless0$, for
three different cases of the orientational distribution of 50 line
defects. In the first and the third cases, Figs. \ref{Fig_Diffusivity}(a)
and \ref{Fig_Diffusivity}(c), the diffusion coefficients $D_{\Vert}$
and $D_{\bot}$ are calculated respectively along and across 50 parallel-oriented
line defects (the distance between lines is different and random).
In the second case, Fig. \ref{Fig_Diffusivity}(b), the lines are
randomly distributed, which results in the isotropic diffusivity,
$D_{{\normalcolor \textrm{rnd}}}\equiv D{}_{xx}\equiv D{}_{yy}$.
As expected, the diffusion coefficients along the lines are higher
than those across the lines, whereas ${\color{black}{\color{black}{\color{black}{\color{red}{\color{black}D_{\bot}}}{\color{black}{\color{red}{\color{black}<}{\color{black}D_{\mathrm{rnd}}}{\color{black}<}}{\color{red}{\color{black}D_{\Vert}}}}}}}$.
After an initial linear increase corresponding to the ballistic regime,
the diffusion coefficients reach their maxima at $t\thickapprox{\color{red}{\color{black}130}}$,
\textcolor{black}{140}, and \textcolor{black}{110} fs for the ``parallel'',
``random'', and ``perpendicular'' cases, respectively. These values
of $D=D_{\max}$ are used to calculate conductivities in Eq. (\ref{Eq_sigmaMax}).
For times $t\gtrsim{\color{red}{\color{black}130}}$, $t\gtrsim{\color{red}{\color{black}140}}$,
and $t\gtrsim{\color{red}{\color{black}110}}$ fs, $D(t)$ decreases
due to the localization effects. Similar temporal behavior of the
diffusivity was established earlier for point\cite{Leconte11,Lherbier12,Radchenko 1}
and line\cite{Radchenko 2} defects in graphene.

\begin{figure*}
\includegraphics[width=0.9\textwidth]{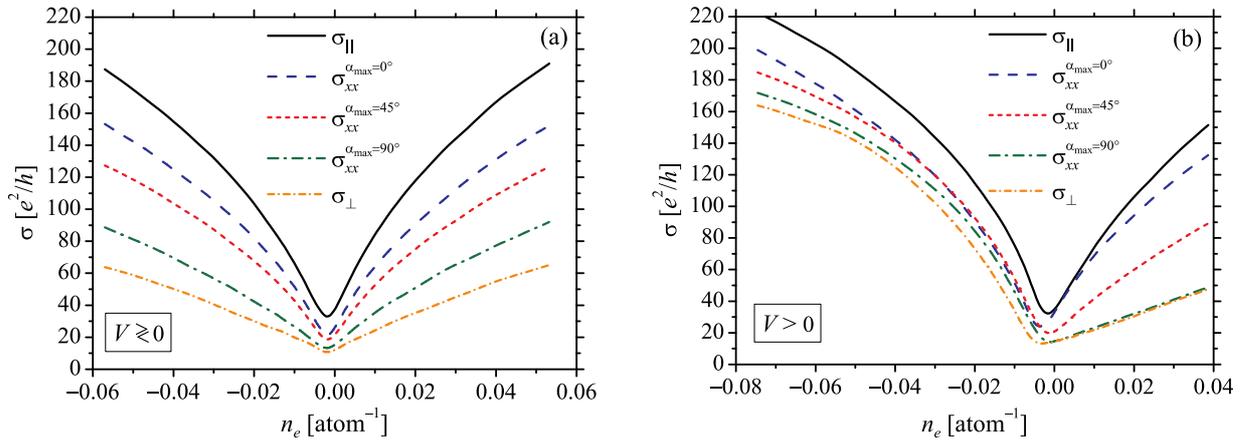}

\caption{(Color online) Conductivities $\sigma_{xx}^{\alpha\mathrm{_{max}}}$
vs. the relative charge carrier density for (a) symmetric, $V\gtrless0$,
and (b) asymmetric, $V>0$, scattering potentials (with $\Delta=0.25u$)
for different configurations of 50 line defects. Each curve averages
50 different defect configurations, corresponding to different distances
between the lines and different prevailing directions in the each
realization. $\sigma_{\Vert}$ and $\sigma_{\bot}$ are the conductivities
in parallel and perpendicular directions to the parallel lines (corresponding
to $\alpha_{\textrm{max}}=0$). }

\label{Fig_Correlation} 
\end{figure*}

\begin{figure*}
\includegraphics[width=0.9\textwidth]{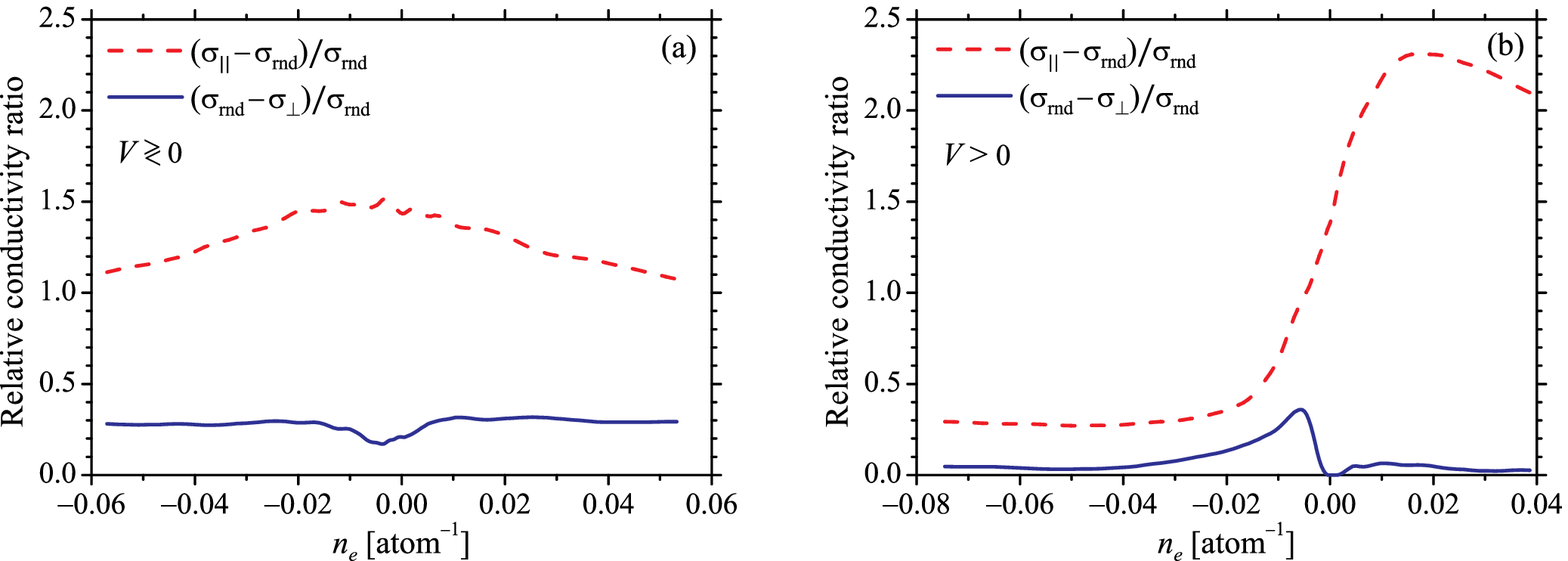}

\caption{(Color online) The ratios $(\sigma_{\parallel}-\sigma_{\mathtt{\mathrm{rnd}}})/\sigma_{\mathtt{\mathrm{rnd}}}$
and $(\sigma_{\mathtt{\mathrm{rnd}}}-\sigma_{\bot})/\sigma_{\mathtt{\mathrm{rnd}}}$
representing relative increase (decrease) of the conductivity in the
direction parallel (perpendicular) to 50 parallel line defects as
compared to the conductivity $\sigma_{\mathtt{\mathrm{rnd}}}=\sigma_{xx}^{\alpha\mathrm{_{max}}=90^{\circ}}$
along the \textit{x}-direction for the same number of randomly distributed
lines.}

\label{Fig_Ratio} 
\end{figure*}

\begin{figure*}
\includegraphics[width=1\textwidth]{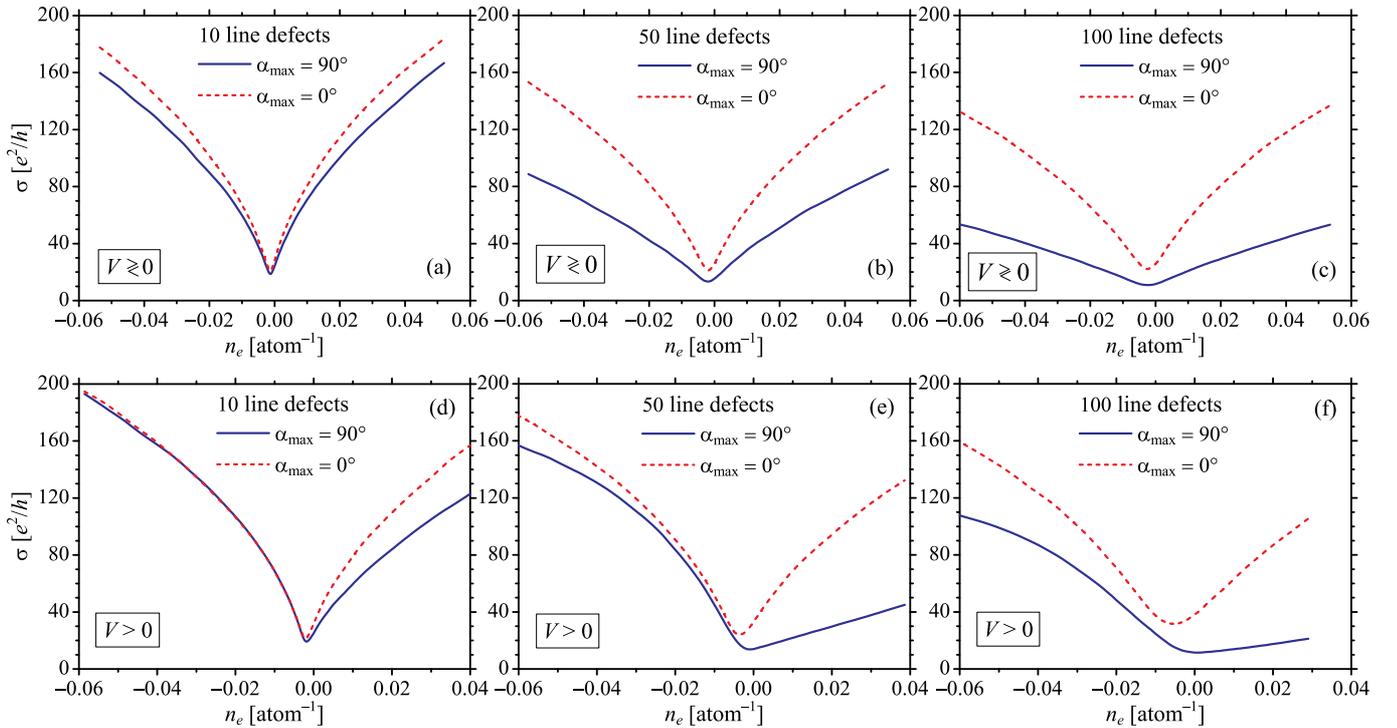}

\caption{(Color online) Conductivities $\sigma_{xx}^{\alpha\mathrm{_{max}}}$
vs. the relative charge carrier density for (a)--(c) symmetric, $V\gtrless0$,
and (d)--(f) asymmetric, $V>0$, scattering potentials (with $\Delta=0.25u$)
for different number (10, 50, 100) of lines for the cases of $\alpha_{\textrm{max}}=90^{\circ}$
and $\alpha_{\textrm{max}}=0^{\circ}$.}

\label{Fig_Cond_10-50-100} 
\end{figure*}

\begin{figure*}
\includegraphics[width=0.9\textwidth]{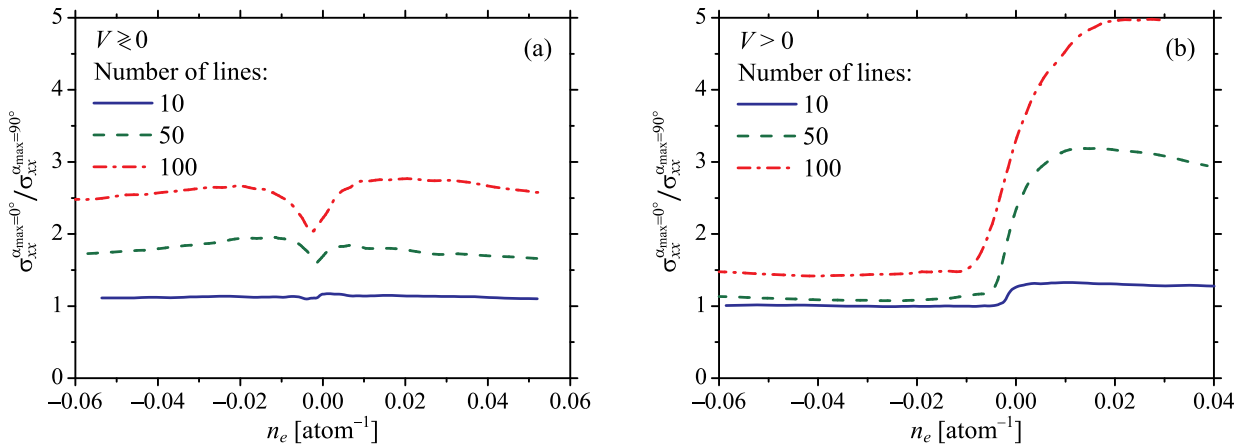}

\caption{(Color online) Enhancement of the conductivity (in terms of the ratio
$\sigma_{xx}^{\alpha\mathrm{_{max}}=0^{\circ}}/\sigma_{xx}^{\alpha\mathrm{_{max}}=90^{\circ}}$)
due to the correlation of line defects for different number (10, 50,
and 100) of lines for (a) symmetric, $V\gtrless0$, and (b) asymmetric,
$V>0$, scattering potentials. }

\label{Fig_Enhancement} 
\end{figure*}

\begin{figure*}
\includegraphics[width=0.93\textwidth]{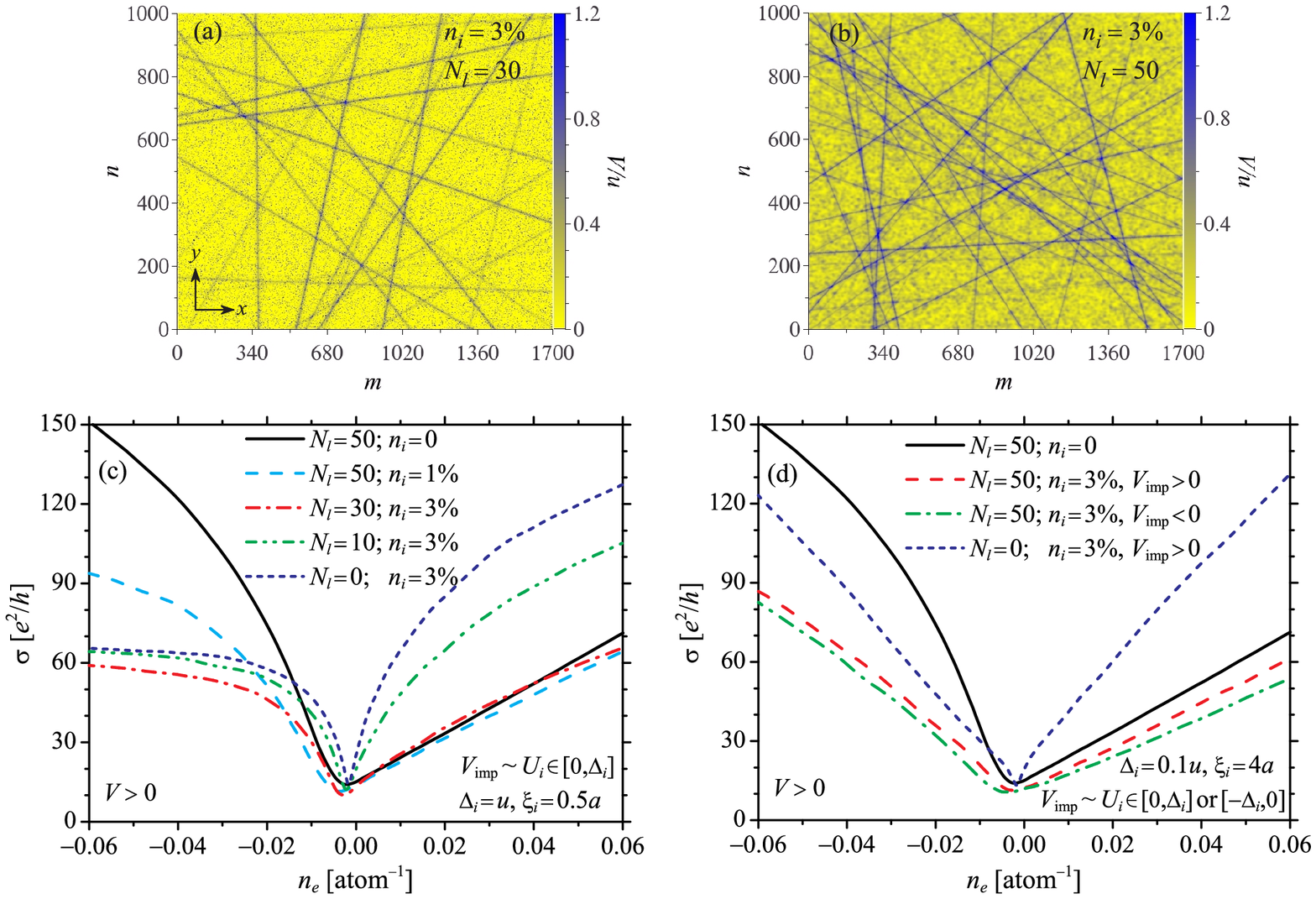}

\caption{(Color online) (a)--(b) Potential distributions in a graphene sheet
of the size $1700\times1000$ sites with (a) $N_{l}=30$ line defects
and $N_{i}=51000$ short-range point scatterers; and (b) $N_{l}=50$
line defects and $N_{i}=51000$ long-range point scatterers. (c)--(d)
The density-dependent conductivity for the graphene sheets with (c)
line defects and short-range point defects, and (d) line defects and
long-range point defects. The line defects are described by the long-range
Lorentzian $V$ (\ref{Eq_Lorentzian}), while the point scatterers
are modeled by the Gaussian $V_{\mathrm{imp}}=\sum_{0}^{N_{i}}U_{i}e^{-\mathbf{r}^{2}/2\xi_{i}^{2}}$
with effective potential radius (a), (c) $\xi_{i}=0.5a$ and (b),
(d) $\xi_{i}=4a$; $U_{i}$ chosen randomly in the range $[0,\Delta_{i}]$. }

\label{Fig_Points+lines} 
\end{figure*}

To ascertain possible effects of the anisotropy, we calculated conductivity
for five different cases of orientations of the lines. Figure \ref{Fig_Correlation}
shows conductivities $\sigma_{\parallel}$ and $\sigma_{\perp}$ for
lines which are parallel ($\parallel$) and perpendicular ($\perp$)
to the \textit{x}-direction, or, in other words, conductivities in
the directions parallel and perpendicular to the parallel-oriented
line defects. Figure \ref{Fig_Correlation} also shows conductivities
along the \textit{x}-direction for line defects with different correlation
angles, $\sigma_{xx}^{\alpha\mathrm{_{max}}=0^{\circ}}$, $\sigma_{xx}^{\alpha\mathrm{_{max}}=45{}^{\circ}}$,
and $\sigma_{xx}^{\alpha\mathrm{_{max}}=90^{\circ}}=\sigma_{\mathrm{rnd}}$
(see Fig. \ref{Fig_Potential} for illustration). The minimum value
$\alpha_{\textrm{max}}=0^{\circ}$ corresponds to the case when all
lines are parallel in each of 50 realizations, but have different
``preferred'' direction in the each configuration. The maximum value
$\alpha\mathrm{_{max}}=90^{\circ}$ corresponds to random (totally
uncorrelated) lines in the each realization. (Note that Fig. \ref{Fig_Potential}
illustrates only one of possible realizations of 1D defects for the
chosen values of $\alpha_{\textrm{max}}$ and the prevailing direction.)

For the case of parallel lines the conductivity along them, $\sigma_{\parallel}$,
substantially exceeds the conductivity in the transverse direction,
$\sigma_{\perp}$, for both symmetric ($V\gtrless0$) and asymmetric
($V>0$) scattering potentials. For instance, for 50 line defects,
the enhancement is up to 3.5 times (see Fig. \ref{Fig_Correlation}).
Such anisotropy is apparently caused by a significantly weaker electron
scattering along the line defects as compared to the transverse direction.
As mentioned in the introduction, the conductance anisotropy has been
revealed in recent experiments,\cite{Kuramochi2012,Yakes2010} and
our calculations are consistent with these results.

Similarly to the behavior of the diffusion coefficient for the case
of randomly oriented lines, the corresponding conductivity, $\sigma_{\mathtt{\mathrm{rnd}}}=\sigma_{xx}^{\alpha\mathrm{_{max}}=90^{\circ}}$,
is also smaller than $\sigma_{\parallel}$ but larger than $\sigma_{\perp}$.
Interestingly, the ratio $(\sigma_{\parallel}-\sigma_{\mathtt{\mathrm{rnd}}})/\sigma_{\mathtt{\mathrm{rnd}}}$
is larger than the ratio $(\sigma_{\mathtt{\mathrm{rnd}}}-\sigma_{\bot})/\sigma_{\mathtt{\mathrm{rnd}}}$
in all the range of electron densities, see Fig. \ref{Fig_Ratio}.
In other words, for the fully random orientation of the line defects,
the corresponding conductivity, $\sigma_{\mathtt{\mathrm{rnd}}}=\sigma_{xx}^{\alpha\mathrm{_{max}}=90^{\circ}}$,
is more close to $\sigma_{\perp}$ than to $\sigma_{\parallel}$.\textcolor{black}{{}
(Note that $|\sigma_{\parallel}-\sigma_{xx}^{\alpha\mathrm{_{max}}=45{}^{\circ}}|\approx|\sigma_{\bot}-\sigma_{xx}^{\alpha\mathrm{_{max}}=45{}^{\circ}}|$}.)
This feature of the conductivity for the case of the random defect
orientation (i.e. $|\sigma_{\parallel}-\sigma_{\mathtt{\mathrm{rnd}}}|\neq|\sigma_{\bot}-\sigma_{\mathtt{\mathrm{rnd}}}|$)
is related to the dominant contribution to the conductivity of the
most strong scatterers,---line defects which are or almost perpendicular
to the given direction of transport. When the correlation angle decreases
from its maximum value ($\alpha_{\textrm{max}}=90^{\circ}$) through
the intermediate one ($\alpha_{\textrm{max}}=45{}^{\circ}$) to the
minimum value ($\alpha_{\textrm{max}}=0^{\circ}$), the conductivity
is also gradually decreases for both symmetric ($V\gtrless0$) and
asymmetric ($V>0$) potentials (Fig. \ref{Fig_Correlation}). Note,
that this behavior of the conductivity for the line defects can be
contrasted with the case of point defects, described by symmetric
(attractive and repulsive) scattering potential, when the correlation
in the defect position practically does not affect the conductivity.\cite{Radchenko 1}

To ascertain how the conductivity of graphene sheets depends on a
concentration of the correlated line defects, we performed numerical
calculations for the same computational domain size (1.7 millions
of atoms), but with different number (10, 50, and 100) of line defects
randomly distributed and parallel-oriented in each of (20) considered
realizations. As Figs. \ref{Fig_Cond_10-50-100} and \ref{Fig_Enhancement}
show, for a given electron density, the relative increase of the conductivity
for the case of fully correlated line defects in comparison to the
case of uncorrelated ones is higher for a larger defect density. For
example, for the case of the symmetric potential with 10 lines, $\sigma_{xx}^{\alpha\mathrm{_{max}}=0^{\circ}}/\sigma_{xx}^{\alpha\mathrm{_{max}}=90^{\circ}}\approx1.15$,
whereas for the case of 100 lines, $\sigma_{xx}^{\alpha\mathrm{_{max}}=0^{\circ}}/\sigma_{xx}^{\alpha\mathrm{_{max}}=90^{\circ}}\approx2.75$
{[}Fig. \ref{Fig_Enhancement}(a){]}; and for the asymmetric potential,
ratios $\sigma_{xx}^{\alpha\mathrm{_{max}}=0^{\circ}}/\sigma_{xx}^{\alpha\mathrm{_{max}}=90^{\circ}}$
may reach values up to $\approx1.35$ and $\approx5$ for 10 and 100
lines, respectively {[}Fig. \ref{Fig_Enhancement}(b){]}. This is
an expected result, because the correlation effect manifests itself
stronger for a larger number of objects-to-be-correlated,---line defects
in our case.

It is noteworthy that the conductivity exhibits a pronounced sublinear
dependence as a function of the electron density. For the case of
uncorrelated line defects such atypical behavior has been attributed
to the extended nature of one-dimensional charged defects.\cite{Radchenko 2}
Apparently, the sublinear density dependence persists also in the
case of correlated extended defects. It should be noted however that
when a defect concentration is increased the sublinear density dependence
of the conductivity gradually transforms into the linear one, see
Fig. \ref{Fig_Cond_10-50-100}. Interestingly that with the increase
of the defect concentration for the asymmetric potential ($V>0$),
$\sigma(n)$ remains sublinear for $n<0$ and transforms into linear
one for $n>0$ (i.e. in the density region where the conductivity
is affected most).

In realistic samples one can expect that the conductivity is also
affected by the presence of point-like scatterers such as adatoms
and charged impurities. \cite{Robinson2008,Wehling2010,Yuan10,Radchenko 1,Radchenko 2,DasSarma_review,Peres_review,Leconte11,Lherbier12}
In the following discussion we focus on the effect of asymmetry of
the conductance for the case of asymmetric potentials ($V>0$). Figure
\ref{Fig_Points+lines} shows the conductivity of graphene sheet for
the case when both extended line defects and point-like defects are
present. The point-like defects are modeled by the Gaussian potential
$V_{\mathrm{imp}}=\sum_{0}^{N_{i}}U_{i}e^{-\mathbf{r}^{2}/2\xi^{2}}$,
where $U_{i}$ is chosen randomly in the range $[0,\Delta]$, and
$\xi$ is the effective potential radius. \cite{Yuan10,Radchenko 1,DasSarma_review}
To model short-range scatteres such as neutral adatoms we choose $\xi=0.5a$
{[}Figs. \ref{Fig_Points+lines}(a) and \ref{Fig_Points+lines}(c){]},
and to model long-range charged impurities we choose $\xi=4a$ {[}Figs.
\ref{Fig_Points+lines}(b) and \ref{Fig_Points+lines}(d){]}.

For further discussion it is important to stress that conductivity
of graphene with point defects with asymmetric potential ($V>0$)
is not symmetric with respect to the Dirac point. (Note that asymmetric
behavior of the conductivity for point-like defects was discussed
in e.g. Refs.~\onlinecite{Wehling2010, Robinson2008}.) However,
this asymmetry is much more pronounced for the case of short-ranged
defects as compared to long-ranged ones {[}cf. short dashed blue curves
in Figs. \ref{Fig_Points+lines}(c) and \ref{Fig_Points+lines}(d){]}.
Let us now consider the case of line defects + short-range point defects
{[}Figs. \ref{Fig_Points+lines}(a) and \ref{Fig_Points+lines}(c){]}.
Each type of scatterers alone exhibits strong asymmetric behavior
of the opposite symmetry {[}cf. black solid and blue short-dashed
curves in Fig. \ref{Fig_Points+lines}(c){]}. Therefore, depending
on the relative concentration of the line- and point defects, the
conductivity can be enhanced whether for positive or negative charge
densities, see Fig. \ref{Fig_Points+lines}(c). In contrast, for the
case of line defects + long-range point defects a character of the
asymmetry of the conductivity is different. In this case, because
of the weak asymmetry of the density dependence of $\sigma$ for the
long-range point defects, the asymmetry of the conductivity of the
graphene is dominated by the asymmetry due to the line defects regardless
of the sign of the scattering potential of point scatterers, see Fig.
\ref{Fig_Points+lines}(d).

\section{Conclusions}

A numerical study of the conductivity of graphene with correlated
extended line defects is performed using the efficient time-dependent
real-space Kubo--Greenwood formalism. The scattering by the line defects
was modeled by an effective long-ranged potential of the Lorentzian
shape. The correlation in the spatial distribution of the line defects
was described by the correlation angle $\alpha_{\textrm{max}}$ (i.e.
the maximum possible angle between any two lines). For the case of
parallel lines ($\alpha_{\textrm{max}}=0$), we find that the conductivity
along the lines, $\sigma_{\parallel}$, substantially exceeds the
conductivity in the transverse direction, $\sigma_{\perp}$, which
agrees well with the experimental measurements for epitaxial graphene.
We found that for a given electron density the relative increase of
the conductivity for the case of the totally correlated line defects
in comparison to the case of uncorrelated ones is higher for a larger
defect density. We also discuss a combined effect of extended line
defects and point short- and long-range defects focusing on the character
of the asymmetry of the conductivity with respect to the Dirac point. 
\begin{acknowledgments}
The authors greatly appreciate discussions with Chariya Virojanadara,
Rositsa Yakimova and Volodymyr Khranovskyy concerning the structure
and type of defects in epitaxial graphene and acknowledge Stephan
Roche for discussion about the time-dependent Kubo--Greenwood approach.
T.M.R. benefited immensely from discussions with Sergei Sharapov and
also expresses gratitude to Valentyn Tatarenko for frequent discussions
on various short- and long-range scattering and interaction potentials.
The authors acknowledge support from the Swedish Institute.\end{acknowledgments}

\end{document}